\documentstyle[11pt,newpasp]{article}

\def\Ha{H$\alpha$}
\def\HI{\ion{H}{\small I}}
\def\CIII{\ion{C}{\small III}}
\def\CIV{\ion{C}{\small IV}}
\def\NV{\ion{N}{\small V}}

\def\OVI{\ion{O}{\small VI}}
\def\OVII{\ion{O}{\small VII}}
\def\OVIII{\ion{O}{\small VIII}}
\def\SII{\ion{S}{\small II}}
\def\SiIII{\ion{Si}{\small III}}

\def\dex#1{10$^{#1}$}
\def\tdex#1{$\times$10$^{#1}$}
\def\kms{km\,s$^{-1}$}
\def\cmm#1{cm$^{-#1}$}
\def\Msun{M$_\odot$}

\def\edcomment#1{\iffalse\marginpar{\raggedright\sl#1\/}\else\relax\fi}
\reversemarginpar

\begin{document}
\title{High-velocity clouds and the Local Group}
\author{B.P. Wakker}
\affil{University of Wisconsin-Madison}

\begin{abstract}
I examine some of the evidence relevant to the idea that high-velocity clouds
(HVCs) are gas clouds distributed throughout the Local Group, as proposed by
Blitz et al.\ (1999) and Braun \& Burton (1999). This model makes several
predictions: a) the clouds have low metallicities; b) there should be no
detectable \Ha\ emission; c) analogues near other galaxies should exist; and d)
many faint HVCs in the region around M\,31 can be found. Low metallicities are
indeed found in several HVCs, although they are also expected in several other
models. \Ha\ emission detected in most HVCs and, when examined more closely,
distant ($D$$>$200~kpc) HVCs should be almost fully ionized, implying that most
HVCs with \HI\ must lie near the Milky Way. No clear extragalactic analogues
have been found, even though the current data appear sensitive enough. The final
prediction (d) has not yet been tested. On balance there appears to be no strong
evidence for neutral gas clouds distributed throughout the Local Group, but
there may be many such clouds within 100 or so kpc from the Milky Way (and
M\,31). On the other hand, some (but not all) of the high-velocity \OVI\
recently discovered may originate in hot gas distributed throughout the Local
Group.
\end{abstract}

\section{Introduction}
The presence of high- and intermediate velocity clouds (IVCs and HVCs; gas with
velocities deviating by more than $\sim$40~\kms\ from differential galactic
rotation) has been known for four decades (Muller et al.\ 1963, Blaauw \&
Tolbert 1966). Some understanding has now been reached about the location and
properties of many of these objects. The IVCs appear to have solar metallicity,
z-heights of $\sim$1 kpc, and possibly represent the return flow of a
Galactic-Fountain type circulation (see Wakker 2001 and references therein). The
Magellanic Stream is a tidal stream extracted from the Small (and maybe also the
Large) Magellanic Cloud (Gardiner \& Noguchi 1996). One HVC (complex~A) lies in
the upper Galactic Halo ($z$=4.6--6.8~kpc; van Woerden et al.\ 1999). For
several other HVCs the intensity of the detected \Ha\ emission also suggests
similar z-heights (Weiner et al.\ 2001), though accurate distances remain
unknown. HVC complex~C clearly consists of gas that has never been part of the
Milky Way before -- it has low metallicity ($\sim$0.1 times solar; Wakker et
al.\ 1999, Richter et al.\ 2001).
\par Although much progress has been made in the understanding of the HVCs, many
questions remain. In particular, there is no consensus about the suggestion that
many of the small HVCs represent the neutral gaseous component of dark matter
halos distributed throughout the Local Group (Blitz et al.\ 1999). I will review
this model (Sect.~2) and the evidence for and against it (Sect.~3). In Sects.~4
and 5, I will discuss the discovery of high-velocity \OVI\ absorption and its
relevance to the connection between HVCs and the Local Group.

\section{\HI\ HVCs and the Local Group}
Fairly early in the study of HVCs, it was suggested that they represented
protogalaxies with distances on the order of 400~kpc (Verschuur 1969). Based on
this suggestion, but using much more data Hulsbosch (1975) calculated the
distances implied if the HVCs in his sample were gravitionally bound objects in
virial equilibrium:
\begin{equation}
D_{vir} = f {6 \alpha \sigma^2 \over 0.236 S G}\  {\rm kpc},
\end{equation}
where $G$ is Newton's constant of gravity, $S$ is the integrated flux in
Jy\,kms, 0.236 is a conversion factor to convert flux to mass, $\sigma$ is the
observed velocity dispersion, $\alpha$ is the angular radius of the cloud, and
$f$ is the ratio $M$(\HI)/$M$(total). Hulsbosch (1975) assumed $f$=1, and thus
found that the implied ``virial distances'' were typically 0.5--10~Mpc. He
concluded that it was unlikely that the HVCs were gravitationally bound and
distant.
\par Verschuur's idea was revived by Blitz et al.\ (1999), who suggested that
HVCs represent the neutral baryonic material in a ten times more massive dark
matter halo. That is, Blitz et al.\ (1999) assumed $f$=0.1. Using the catalogue
of Wakker \& van Woerden (1991) they derived that the median distance of a
virially stable HVC was reduced to $\sim$1~Mpc, and the median implied mass was
$\sim$3\tdex7~\Msun, with an integrated total mass in the HVCs of
\dex{10}~\Msun. They combined these numbers with a basic model of the evolution
of the Local Group that predicted the present-day locations and velocities of
test particles.
\par Although Blitz et al.\ (1999) concluded that the predicted and observed
distributions were similar, this conclusion has remained controversial. A major
contribution of the Blitz et al.\ (1999) article was their effort to make
testable predictions. These included: a) the HVCs should have subsolar
metallicities; b) they should have undetectable \Ha\ emission; c) analogues
should be seen in other groups and/or near other galaxies; d) with higher
sensitivity many faint HVCs should be seen in the region around M\,31. In the
next section I will review the current status of these tests.
\par Following Blitz et al.\ (1999), Braun \& Burton (1999) suggested that only
the small HVCs (which they termed ``compact HVCs'' or CHVCs'') were distant
Local Group objects. They followed this up by several studies, culminating in
the series of papers by de Heij et al.\ (2002a,b,c). In the final version of
their model, there were originally some 6000 CHVCs, but most are unstable
against disruption by tides or ram pressure. About 1000 survive, but most are
too faint to be detected, or have the wrong velocity to be considered an HVC,
leaving just $\sim$150 CHVCs at present. Most lie within 300~kpc of either M\,31
or the Milky Way. The total \HI\ mass in these clouds is about \dex9~\Msun,
while the most massive CHVC has M$\sim$\dex7~\Msun. Thus, in this model, the
CHVCs are about 10 times less massive and 5 times closer than in the Blitz et
al.\ (1999) model.

\section{Observational tests}
\subsection{Metallicities}
Local Group HVCs should have subsolar metallicities. Reliable values are known
for just two HVCs -- complex~C and the Magellanic Stream. Fox et al.\ (2004)
summarize the determinations of S/H and O/H in complex~C made by Wakker et al.\
(1999), Gibson et al.\ (2001), Richter et al.\ (2001), Collins et al.\ (2003)
and Tripp et al.\ (2003), bringing them on the same solar abundances scale.
(O/H) is 0.14--0.19 times solar, while (\SII/\HI) varies from 0.09 to 0.46 times
solar, with a strong dependence on N(\HI) that is clearly due to varying
ionization. For the Magellanic Stream Lu et al.\ (1998) and Gibson et al.\
(2000) find (SII/HI)= 0.25--0.30 times solar, equal to the value expected for
Magellanic gas.
\par A preliminary analysis of new data from the Far Ultraviolet Spectroscopic
Explorer (FUSE) shows that HVC complex~A ($l$$\sim$150\deg, $b$$\sim$35\deg,
$v$$\sim$$-$150~\kms) probably has Z$\sim$0.1 solar, complex~WB
($l$$\sim$240\deg, $b$$\sim$20\deg, $v$$\sim$+100~\kms) has Z$\sim$1 solar,
complex~WD ($l$$\sim$280\deg, $b$$\sim$+25\deg, $v$$\sim$+100~\kms) has
Z$\sim$0.1 solar, and cloud WW84 ($l$=125\deg, $b$=+42\deg, $v$=$-$205~\kms) has
Z$\sim$0.05 solar.
\par Clearly, some of the HVCs have strongly subsolar metallicities, showing
that they are not Galactic clouds. For complex~C we also know that the (N/O)
ratio is ten times lower than that in the nearby ISM, and that it has relatively
high D/H (Sembach et al.\ 2004). In the Local Group model, nearby
low-metallicity HVCs like complex~A and C are the gaseous component of a small
halo dark-matter halo that has come close to the Milky Way. However, although
low metallicities are expected in the Local Group model, it is not sufficient
evidence. Complex~C (and other HVCs) could be the remnant of a tidally stripped
dwarf galaxy, whose stars are no longer recognizable. Or they could be
condensations in a hot ($T$$>$\dex6~K) extended ($R$$>$50~kpc), tenuous
($n$$<$\dex{-4}~\cmm3) gaseous Galactic corona.

\subsection{\Ha\ observations}
Distant HVCs will only be ionized by the extra-galactic radiation field, which
is insufficient to produce observable \Ha\ emission. However, \Ha\ emission is
detected from many HVCs: Weiner \& Williams (1996; the Magellanic Stream), Tufte
et al.\ (1998; complexes~A, C and M), Bland-Hawthorn et al.\ (1998; complex~GP),
Tufte et al.\ (2002; four CHVCs), Weiner (2003; complexes~L, AC, GCN). Typical
intensities lie in the range 0.1--0.3~Rayleigh, with a few limits
$<$0.1~Rayleigh. The results of these now multitudinous detections of \Ha\
emission from HVCs imply that most of them must lie no more than several tens of
kpc from the Milky Way. However, there are still many candidate Local Group HVCs
whose \Ha\ emission has not yet been observed, so the case is not closed.
\par Maloney \& Putman (2003) looked at this problem from the theoretical side.
For CHVCs at distances of $\sim$1~Mpc, the implied \HI\ volume densities are
$\sim$3\tdex{-4}~\cmm3. With N(\HI) a few \dex{19}~\cmm3\ such clouds will be
almost fully ionized by the extragalactic ionizing radiation field (5\% neutral
fraction). If the CHVCs also have a dark matter to baryon ratio of 10, the
observed amounts of \HI\ then imply that CHVCs should dominate the dynamics of
the Local Group, and would have line widths much larger than is observed. If the
CHVCs lie in the Galactic halo ($D$$<$200~kpc) their properties give consistent
results.

\subsection{HVCs near other galaxies and galaxy groups}
Blitz et al.\ (1999) argued that searches for HVC analogues near other galaxies
and in other galaxy groups were still rather limited and not quite sensitive
enough. De Heij et al.\ (2002c) make the specific prediction that there are 95
CHVCs with mass larger than 3\tdex6~\Msun, and 45 with mass $>$5\tdex6~\Msun,
but just 1 with M$\sim$\dex7~\Msun. If one were to observe the Local Group from
a distance, these would be distributed over an area of some 1.5$\times$1~Mpc.
\par Since 1999 much data has appeared, with detection limits that should have
been sufficient to find clouds with masses of \dex6~\Msun. But no free-floating
starless clouds have been found. For instance, Pisano et al.\ (2004) mapped part
of two nearby galaxy groups with the VLA and the DRAO telescope with
$\sim$1~arcmin ($\sim$10~kpc) resolution. The groups (GH\,144 and GH\,158, at 33
and 45~Mpc distance, with group radii of $\sim$1.5~Mpc) were found in the
catalogue of Geller \& Huchra (1983). They are fairly similar to the Local
Group. GH\,144 has two large spirals and one large elliptical, while there are 4
large spirals and 1 large elliptical in GH\,158. Toward the centers of the
pointings, the VLA data had a detection limit of 1--1.5\tdex6~\Msun, or
$\sim$\dex{18}~\cmm2. At the group's distances the VLA primary beam corresponds
to $\sim$400~kpc. The DRAO data were a factor $\sim$20 less sensitive, but
covered a 5 times larger area. For both groups, the VLA fields were centered on
an UV-bright AGN, and pass about 450~kpc from the nearest group galaxy. Since
these pointings cover $\sim$5--10\% of the group's area, the de Heij et al.\
(2002c) model predicts that on the order of 5--10 \HI\ clouds should have been
easily detectable. Two \HI\ objects were detected in GH\,144. However,
inspection of the Digital Sky Survey shows the presence of a faint dwarf galaxy
near both of these (see Fig.~1).

\begin{figure}[t]
\plotfiddle{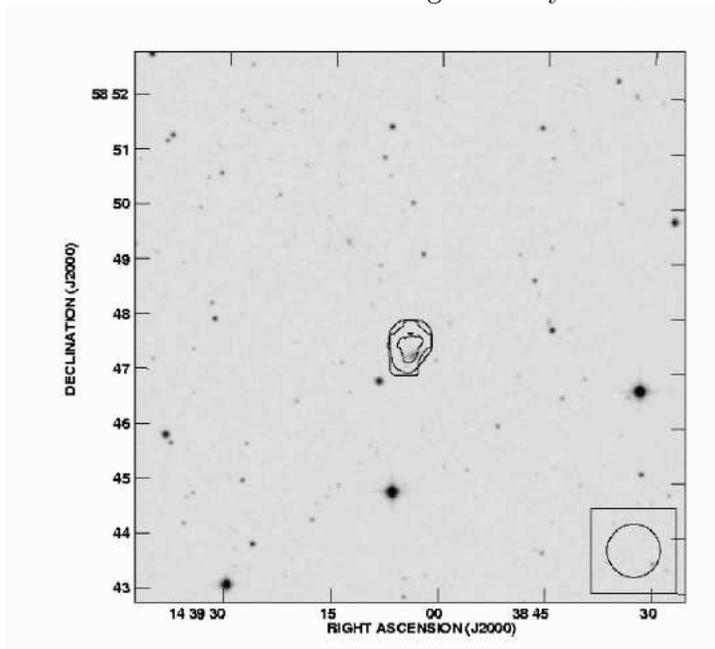}{7.5cm}{0}{50}{50}{-150}{-80}
\caption{\HI\ column density overlaid on the blue DSS-2 image of PWWF
J1439+5847. \HI\ contours are at 0.5, 1, 2, 3\tdex{19}~\cmm2.}
\end{figure}

\par Another deep survey of Local Group analogues was done by Zwaan (2001). He
did a total of 300 Arecibo pointings on 5 galaxy groups similar to the Local
Group, with a detection limit of 7\tdex6~\Msun. All detections could be
attributed to optical galaxies. From a statistical analysis, he concluded that
if $f$=0.1 there is room for at most $\sim$10 clouds with
M(\HI)$>$3\tdex7~\Msun. However, if $f$=0.02 there could be 500 or more HVCs in
the Local Group. A value of $f$=0.02 can be achieved if the HVCs consist of 20\%
\HI\ and 80\% H$^+$, but it also implies that virially stable HVCs must be
closer than 200~kpc on average.
\par A slightly different approach was taken by Miller et al.\ (2003), who made
deep (detection limit $\sim$5\tdex5~\Msun) \HI\ maps of a
$\sim$100$\times$100~kpc region around M\,51 and M\,83. M\,51 has undergone a
tidal interaction with NGC\,5195 and lots of debris clouds are found around it
(see Fig.~2). M\,83 on the other hand is isolated and no HVCs are found away
from the main disk, although there is gas with anomalous velocities up to
100~\kms\ projected on the disk of M\,83.

\begin{figure}[t]
\plotfiddle{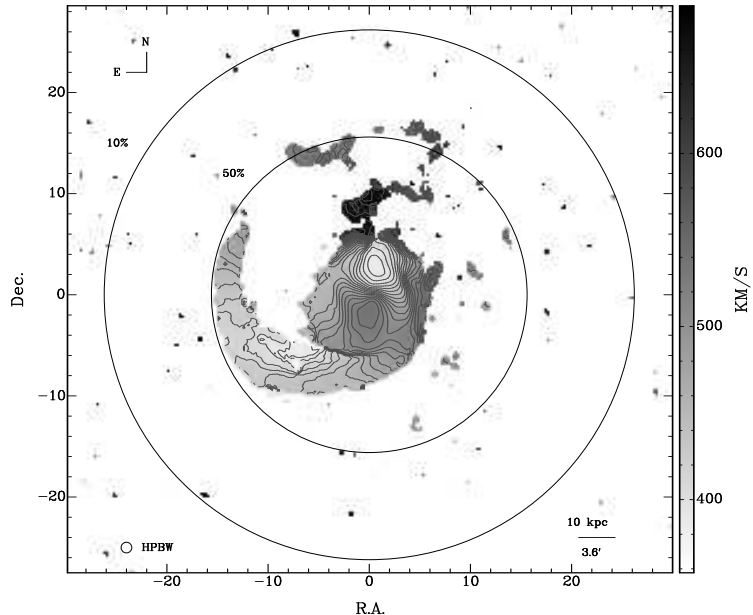}{7.5cm}{0}{58}{58}{-150}{-55}
\caption{\HI\ velocity field of the M\,51/NGC\,5195 system, showing the many
HVCs surrounding this galaxy (from E.\ Miller, U.\ Mich.\ thesis).}
\end{figure}

\par In summary, although the low metallicities found for several HVCs are
consistent with the Local Group hypothesis, the other tests proposed by Blitz et
al.\ (1999) seem to fail. HVCs {\it are} usually detected in \Ha. \dex7~\Msun\
analogues are {\it not} found in other galaxy groups. Theoretical considerations
about their state of ionization show that to get the observed \HI\ column
densities, the ensemble of distant and massive HVCs would be more massive than
the Milky Way and M\,31 combined. On the other hand, the limits set from \Ha\
and from observing other galaxy groups still allow for a population of clouds
within $\sim$200~kpc from the Milky Way. The comparison with the M\,51/M\,83
\HI\ maps suggests that if such a population exists, it might be related to
tidal effects.

\section{Highly ionized HVCs}
A new angle on the possible relationship between HVCs and the Local Group has
been provided by \OVI\ absorption line data obtained using the Far Ultraviolet
Spectroscopic Explorer (FUSE). This satellite was launched on 24 June 1999, and
takes spectra between 912 and 1187~\AA, with a velocity resolution of
$\sim$20~\kms\ (Moos et al.\ 2000; Sahnow et al.\ 2000). This wavelength range
includes the lines of O$^{+5}$ at 1031.926 and 1031.617~\AA. O$^{+5}$ is a good
tracer of gas with temperatures of a few \dex5~K, as it is difficult to create
by photo-ionization. A survey of the \OVI\ absorption toward 100 UV-bright
extragalactic targets was completed in 2003 (Wakker et al.\ 2003; Savage et al.\
2003, Sembach et al.\ 2003). In 59 of the sightlines a total of 84 high-velocity
\OVI\ components were detected, with velocities $|v_{\rm LSR}|$ between 100 and
450~\kms. In the 18 months since the survey cut-off date this sample has grown
to 117 high-velocity \OVI\ components in 83 out of 138 sightlines. Figure 3
shows several examples of \OVI\ spectra containing high-velocity \OVI\
components.
\par Sembach et al.\ (2003) distinguished several different kinds of
high-velocity \OVI. Based on their velocities, six high-velocity \OVI\
components can be identified as absorption in four nearby galaxies (M\,31,
M\,32, M\,33, M\,101). Another three seem to be associated with extended halos
of slightly more distant galaxies. Many are in or near directions where
high-velocity \HI\ is also found at the same velocity, particularly toward
complexes A, C, WD, the Outer Arm and the Magellanic Stream. A detailed analysis
of the available \OVI, \NV\ and \CIV\ absorption for the complex~C sightlines by
Fox et al.\ (2004) suggests that the most likely explanation for the presence of
hot gas associated with complex~C is that this HVC is embedded in a
$T$$>$\dex6~K hot gaseous corona. A similar origin seems likely for the other
associations between \HI\ HVCs and \OVI\ HVCs, although a detailed analysis of
the physical conditions in these high-velocity \OVI\ components is still
lacking. Since the Magellanic Stream is about 50~kpc distant, the extent of the
$T$$>$\dex6~K corona must be at least 60~kpc.
\par Of the remaining sightlines with high-velocity \OVI, four pass over the
Galactic Center. Possibly this indicates the presence of an outflow, but much
more study is needed. For nine high-velocity \OVI\ components no obvious
association with \HI\ HVCs or galaxies can be found.
\par Of the remaining 48 high-velocity \OVI\ features, 24 occur at high negative
velocity, all at $b$$<$0\deg, with $l$ between 15\deg\ and 140\deg. Another 24
occur at high positive velocity, all at $b$$>$0\deg, with $l$ between 170\deg\
and 310\deg. There are several possible explanations for these absorption
components.
\par A) Since about 2/3rds of the high-positive velocity \OVI\ components show
up as an extended wing on the lower-velocity Milky Way absorption, it is
possible that these absorptions originate in outflowing hot gas, associated with
a Galactic Fountain flow. This does not explain the high-negative velocities
seen in the southern sky, however.


\begin{figure}[t]
\plotfiddle{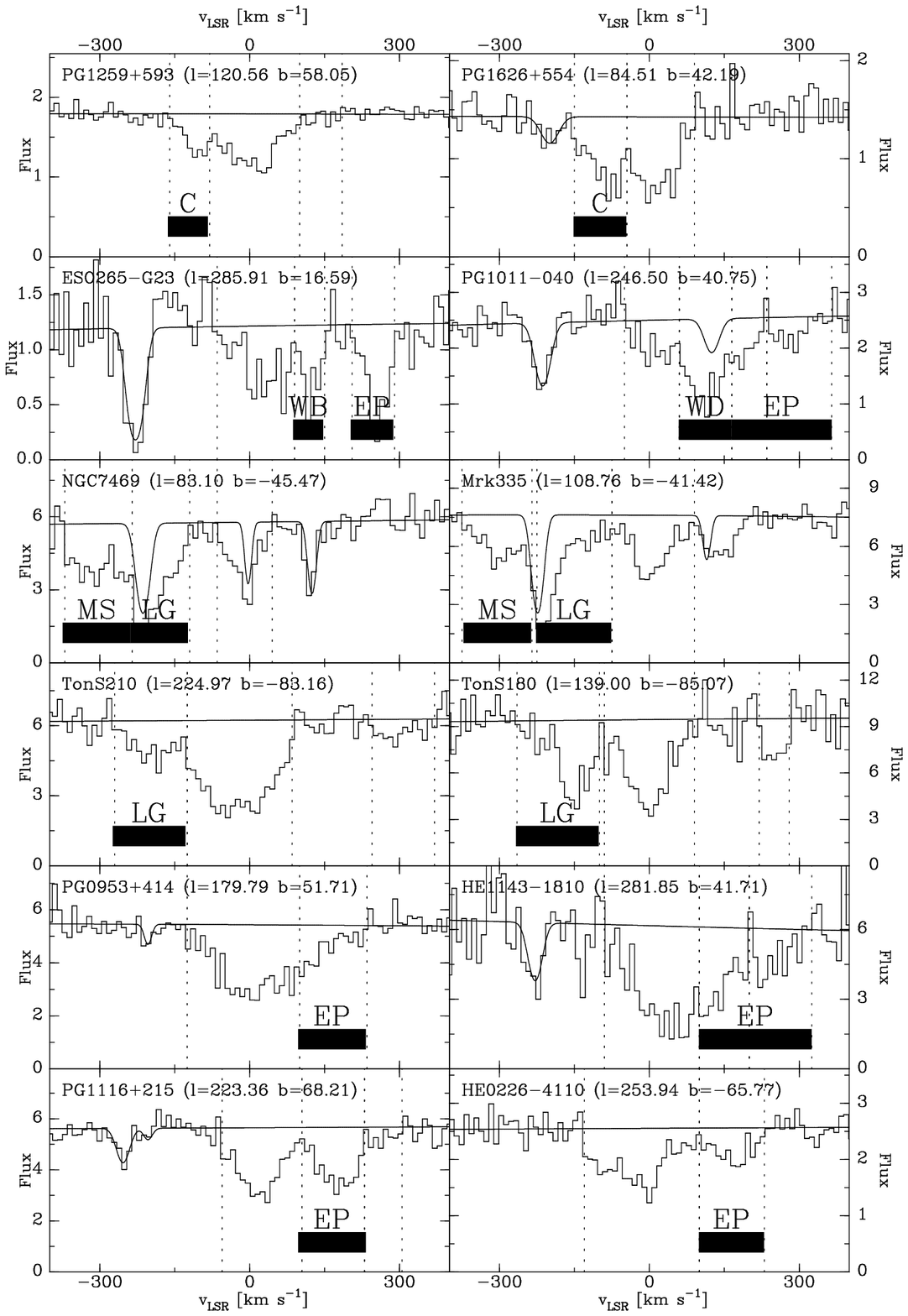}{10.8cm}{0}{90}{90}{-225}{-320}
\end{figure}

\par B) The velocity pattern of these components also reflects that of the
Magellanic Stream. However, in the southern sky the \HI\ velocities (as well as
the predicted velocities for Stream gas) are much more negative than the
velocities of these \OVI\ absorptions. In fact, in many southern directions two
high-velocity \OVI\ components are seen, one of which has velocities similar to
those expected

\ \par
\vskip6.3cm\par
{Figure 3. Examples of high-velocity \OVI\ seen in FUSE spectra. The dotted
vertical lines indicate the identified components. The thick black bars identify
the high-velocity \OVI: ``C/WB/WD''= complex~C/WB/WD. ``MS''= Magellanic Stream,
``LG''= possible Local Group gas, ``EP''= extreme-positive velocity HVC.}

\newpage\noindent
and observed for the Stream. In the northern sky this separation is
not as clear. The high-positive velocity \HI\ in the Leading Arm of the
Magellanic Stream extends to $b$$\sim$30\deg, where only three sightlines are
known. However, many (though not all) of the high-positive velocity \OVI\ lies
along the expected orbit of Stream gas.
\par C) It is notable that most Local Group galaxies in the southern sky have
negative velocities, while in the northern sky most Local Group galaxies are
observed at positive velocities. Therefore, the high-velocity \OVI\ may
originate in a hot gaseous filament extending between the Milky Way and M\,31,
through which the Milky Way is moving. Such an explanation fits very well with
current hydrodynamical simulations of the universe (Cen \& Ostriker 1999, Dav\'e
et al.\ 2001).
\par It remains unclear which of these three explanations is correct. More
analysis is necessary.

\section{Hot Local Group gas?}
Observations relevant to this question are provided by the detection of \OVII\
and \OVIII\ absorption at X-ray wavelengths toward three targets (Nicastro et
al.\ 2002, Fang et al.\ 2003, Rasmussen et al.\ 2003). Nicastro et al.\ (2003)
claim that the high-velocity \OVI\ absorptions have Local-Group like kinematics,
as well as that the \OVI, \OVII, and \OVIII\ absorption occur in the same
\dex{6-7}~K hot gas. According to them the \OVII, and \OVIII\ must originate in
very tenuous gas, implying a Mpc size scale. We now consider these claims in
more detail.
\par We have done a more thorough analysis of the kinematics of HVCs (Wakker et
al., in preparation), and show that the HVC kinematics indicate that they form a
non-rotating population. Beyond that, it is not possible to determine with
confidence whether they are concentrated within 200~kpc from the Milky Way or
whether they are more distant. Blitz et al.\ (1999) and de Heij et al.\ (2002c)
compare the average velocity and the velocity dispersion of the HVC sample to
that of models. Nicastro et al.\ (2003) calculate an average velocity vector by
decomposing the observed radial velocities into three spatial components and
then averaging those. We show that the derived velocities (especially the
average velocity of the cloud ensemble) strongly depend on the selection
criteria used to define the HVC sample, making any detailed comparison between
data and model relatively unreliable.
\par \OVII\ and \OVIII\ have been detected toward several objects with Chandra
(resolution 660~\kms) and XMM (resolution 300~\kms). In all cases the lines are
not resolved, and a $b$-value must be assumed to derive columm densities. If
collisional ionization equilibrium is assumed, then the ratio N(\OVII)/N(\OVI)=
250--450 for $T$ in the range \dex{6-7}~K, while N(\OVIII)/N(\OVI) is a much
more sensitive temperature indicator.
\par Fang et al.\ (2003) measure N(\OVII)=1.8$^{+0.2}_{-0.7}$\tdex{16}\cmm2\ in
the direction of 3C\,273, using Chandra. They assume that $b$$>$100\kms, which
is about the extent of the \OVI\ high-velocity wing. They further argue that $b$
must be large because other \OVII\ transitions are not detected. Rasmussen et
al.\ (2003) derive N(\OVII)$>$0.8\tdex{16}\cmm2, using XMM. The \OVI\ HVC in
this direction has N(\OVI)=3.9\tdex{13}\cmm2, implying \OVII/\OVI$\sim$200--450.
If both originate in the same gas, this ratio implies $T$=\dex{6-7}~K. For a
pathlength $L$=1~Mpc, N(\OVII) implies a density $n = {N\over L A({\rm O}) f(O
{\small VII})}$. For log $A$(O)=$-$4.24 (0.1 solar) and $T$=\dex{6-7}~K this
gives the rather high value of $n$$\sim$\dex{-3}--1~\cmm3, showing that the
\OVI\ and \OVII\ probably do not originate in a homogeneous Local Group medium
that is in collisional ionization equilibrium.
\par Nicastro et al.\ (2002) observed PKS\,2155$-$304 with Chandra and derive
N(\OVII)=4$^{+2.6}_{-2.1}$\tdex{15}\cmm2\ and
N(\OVIII)=5.2$^{+4.3}_{-3.9}$\tdex{15}\cmm2. From XMM Rasmussen et al.\ (2003)
find N(\OVII)$>$4.5\tdex{15}\cmm2\ toward PKS\,2155$-$304. This assumes
$b$=200\kms, which is the total width of the \OVI\ absorption in this direction.
However, if the \OVII\ is only associated with the high-velocity \OVI,
$b$$\sim$100\kms, and the implied N(\OVII) is a factor 2 higher. Since N(\OVI,
HVC)=1.1\tdex{14}\cmm2, \OVII/\OVI=35--70 and OVIII/\OVI=50--100. As Nicastro et
al.\ (2002) note, these ratios imply that the \OVI, \OVII\ and \OVIII\ cannot
originate in a single absorber at constant temperature. They then go on to make
photo-ionization models and conclude that these can explain the ionic ratios if
the gas is tenuous, and thus has a large pathlength. However, the implied
parameters are inconsistent with those derived for the same HVC based on the
measured \CIV\ and \CIII\ column densities (Sembach et al.\ 1999). Clearly more
analysis is needed, and the location of this absorber remains uncertain.
\par Finally, with XMM Rasmussen et al.\ (2003) detect \OVII\ and \OVIII\ toward
Mrk\,421, with N(\OVII)$>$4.8\tdex{15}\cmm2. The \OVI\ HVC in this direction has
N(\OVI)=2.3\tdex{13}\cmm2, giving a ratio of $\sim$200, compatible with
$T$=\dex6~K.
\par Clearly, it is possible that some of the high-velocity \OVI\ originates in
the same hot gas that also produces \OVII\ and \OVIII\ absorption. However, the
evidence for this remains circumstantial and a more complete analysis that
includes the other ions (\NV, \CIV, \CIII, \SiIII, etc.) is required. Also, the
assumption of collisional ionization equilibrium is easily violated.
Pure photo-ionization models may or may not be more appropriate. What is clearly
needed are resolved \OVII\ and \OVIII\ profiles.

\end{document}